\documentclass[12pt, draftclsnofoot, onecolumn]{IEEEtran}
\usepackage{fancyhdr}
\usepackage{authblk}
\usepackage{amsmath}
\usepackage{subfigure}
\usepackage{setspace}
\usepackage{soul,cite}
\usepackage{extarrows}
\usepackage{color}
\usepackage{mathrsfs}
\usepackage{cite}
\usepackage{graphicx}
\usepackage[justification=centering]{caption}
\usepackage{booktabs}
\usepackage{algorithmic} 
\usepackage{amssymb}
\usepackage{amsthm}
\usepackage{bm}
\usepackage{enumerate}
\usepackage{stfloats}
\usepackage{mathtools}
\usepackage{caption}
\usepackage{makecell,multirow,diagbox}
\usepackage{tikz}
\usepackage[marginal]{footmisc}

\usepackage{pgfplots}
\pgfplotsset{compat=newest}
\usetikzlibrary{plotmarks}
\usetikzlibrary{arrows.meta}
\usepgfplotslibrary{patchplots}
\usepackage{grffile}
\newtheorem{theorem}{Theorem}
\begin{document}
\title{
\vspace{-3mm}
\huge{A New High Energy Efficiency Scheme Based on Two-Dimension Resource Blocks in Wireless Communication Systems}}

\author[*]{Kang Liu}
\author[*,+]{Zaichen Zhang(Corresponding author)}
\author[*,+]{Jian Dang}
\author[*,+]{Liang Wu}
\author[*]{Bingchen Zhu}
\author[*]{Lei Wang}
\author[*,+]{Chuan Zhang}
\affil[*]{Southeast University}
\affil[+]{Purple Mountain Laboratories, Nanjing 211111, China}

\maketitle

\footnote{This work is supported by NSFC projects (61960206005, 61971136, 61803211, and 61871111), Jiangsu NSF projects (BK20191261, BK20200820, and BK20200393), the Fundamental Research Funds for the Central Universities (2242021k30043 and 2242021k30053), Research Fund of National Mobile Communications Research Laboratory, and Zhishan Youth Scholar Program of SEU.}

\begin{abstract}
Energy efficiency (EE) plays a key role in future wireless communication network and it is easily to achieve high EE performance in low signal-to-noise (SNR) regime. In this paper, a new high EE scheme is proposed for a multiple-input–multiple-output(MIMO) wireless communication system working in the low-SNR regime by using two-dimension resource allocation. First, we define the high EE area based on the relationship between the transmission power and the SNR. To meet the constraint of the high EE area, both frequency and space dimension are needed. Besides analysing them separately, we decided to consider frequency and space dimensions as a unit and proposed a two-dimension scheme. Furthermore, considering communication in the high EE area may cause decline of the communication quality, we add quality-of-service(QoS) constraint into the consideration and derive the corresponding EE performance based on the effective capacity. We also derive an approximate expression to simplify the complex EE performance. Finally, our numerical results demonstrate the effectiveness of the proposed scheme.
\end{abstract}

\begin{IEEEkeywords}
Energy efficiency, high EE area, two-dimension scheme, quality-of-service(QoS), low-SNR approximation
\vspace{-1mm}
\end{IEEEkeywords}
\IEEEpeerreviewmaketitle

\doublespacing
\section{Introduction}
\IEEEPARstart{W}ith the development of five-generation (5G)\textsuperscript{\cite{boccardi2014five}}, the transmission rate of wireless network is rapidly increasing and the link capacity almost reaches the Shannon capacity limit. While the rate growing, the corresponding system’s power consumption also increases dramatically. It is shown that the total energy used by the communications industry (CI) has taken up more than 5\% of the worldwide electric energy consumption and for the worst-case scenario, that CI electricity usage contributes up to 23\% of the globally released greenhouse gas emissions in 2030\textsuperscript{\cite{andrae2015global,coroiu2019energy}}. Therefore, methods to improve system EE performance become a severe issue in future communication system.

Energy efficiency (EE) is defined as the amount of information that can be delivered per unit energy. The basic relationship between information and energy is given by Shannon's information theory. Shannon capacity indicates that maximum data transmission rate is related to the transmitted power and the transmission bandwidth. Based on that, an effective way is to increase transmission bandwidth\textsuperscript{\cite{cook1982special,gardner1979fading,verdu1990channel,gallager2002power}}. Recently, wideband communication has been regarded as an important part to achieve high EE for communication system. Verdú in \cite{verdu2002spectral} determined the minimum bit energy required for reliable communication over a general class of channels by considering the Shannon capacity formulation, and studied of the spectral efficiency–bit energy tradeoff in the wideband regime. Also, the work in \cite{miao2013energy} studied the detailed energy efficiency performance in wideband regime, which is characterized by wide signal bandwidth and low spectral efficiency. They also considered circuit power into the power consumption for system’s EE performance and discussed the globally optimal energy efficient design for an end-to-end communications system. Quality of service (QoS)\textsuperscript{\cite{tang2007quality}} requirement is also of paramount importance especially when system's signal-to-noise (SNR) is pretty low. In \cite{qiao2011energy}, energy efficiency is considered in the presence of channel uncertainty and QoS limitations in the form of queueing constraints. By employing the effective capacity formulation, the spectral efficiency–bit energy tradeoff in the low-power and wideband regimes. Energy efficiency in wideband regime in the presence of QoS constraints has been studied in \cite{gursoy2009analysis}. The analysis has been conducted under different assumptions on the channel side information (CSI) and provided a characterization of the energy-bandwidth-delay tradeoff.

Besides working in a wideband regime, another effective way to improve system’s EE is to use multiple-input multiple-output (MIMO) technology\textsuperscript{\cite{dahlman20103g,marzetta2010noncooperative}}. MIMO has been applied to many wireless standards since it can significantly improve the capacity and reliability of communication\textsuperscript{\cite{gesbert2008single,caire2003achievable,viswanath2003sum,vishwanath2003duality}}. And comparing with single-antenna network, MIMO system’s energy efficiency can be much higher too\textsuperscript{\cite{lu2014overview,liu2015massive}}. EE in MIMO systems have been studied in \cite{bjornson2017massive,ngo2013energy,bjornson2015optimal}. In \cite{bjornson2017massive}, the energy efficiency of a massive MIMO system has been analyzed. Considering circuit power consumption, the work in \cite{ngo2013energy} has shown the relationship between the number of antennas and the transmitted power work during system operation. With spatially correlation, EE of a massive MIMO system has been analyzed in \cite{bjornson2015optimal}. Based on the analytical results, an optimal power allocated scheme has been developed. There also has been plenty researches on wideband MIMO systems\textsuperscript{\cite{miao2009cross}}. For example, in \cite{liu2003capacity}, both narrow-band and wide-band channels have been considered in realistic correlated MIMO channels and scaling behavior of ergodic capacity has been studied as a function of both the number of antennas and bandwidth in a Rayleigh-fading environment. In \cite{do2013optimal}, a transmit strategy has been considered in a MIMO broadcast channel in the wideband regime in order to maximize the EE performance.

In the past few years, a number of surveys and tutorials have been contributed to the EE improvements. In particular, a basic EE model has been introduced in \cite{li2011energy} and advanced techniques for energy efficiency are summarized, including orthogonal frequency division multiple access (OFDMA), multiple input multiple output (MIMO), and relay networks. In \cite{feng2012survey}, a complete EE model has been presented and a dedicated discussion on the energy efficient wireless communications has been characterized. With circuit power into consideration, a fundamental tradeoff between spectrum efficiency(SE) and EE has been analyzed in \cite{chen2011fundamental,zhang2019first}. In \cite{zhang2016fundamental}, a comprehensive overview on the extensive on-going research efforts has been provided and several research progresses have been discussed.

According to the research works, communication with high EE always lead to low transmission power, which will cause the decreasing of the system capacity. Using resources on frequency dimension is a reliable solution to this problem. Also, MIMO plays an important role in improving EE performance. If we combine them as a unit in a point-to-point MIMO system and dynamically adjust the channel resources on two dimensions, we can achieve both high EE and high system capacity.

In this paper, we focus on the high EE transmission in a point-to-point MIMO system. This paper can be divided into two parts. In the first part, we present a reliable scheme to make the system's EE stay at high level and meanwhile meet capacity requirements. Based on the relationship between EE performance and SNR, we find a high EE area. Then we demonstrate the EE in spacial and frequency dimension and find equivalence between them. Based on that, we further decide to consider them as a unit and present a two-dimension scheme. In the second part, we present the system model and derive the EE model for the system in the high EE area when QoS constraint and effective capacity are under consideration. Using the low SNR condition, we further derive an approximate result for the system's EE. More specifically, the contributions of the paper are the following:
\begin{enumerate}[i)]
\item We come up with the idea with the needed SNR to communicate in the high-EE area. Based on it, we further develop a two-dimension resource allocation scheme to put both spacial and frequency dimensions as a unit.
\item Using our proposed scheme, we determine the EE model for the system working in the high EE area considering QoS constraint by employing the effective capacity formulation and specific QoS exponent.
\item Using the low SNR condition for the high EE area, we simplify the EE expression derived before and get an approximate result.
\end{enumerate}

The paper is structured as follows. In Section II, we define the high EE area, analyze the equivalency for the EE performance of frequency dimension and space dimension, respectively, and present a two-dimension resource blocks allocation scheme. In section III, we introduce the energy efficiency model under QoS constraint and derive corresponding EE performance for the system working in the high EE area. We also use low SNR condition to achieve an approximate EE expression. In Section IV, we present numerical result for the scheme. Finally, the paper is concluded in Section V.

\section{Equivalency of Frequency and Space Dimensions}
Consider a point-to-point communication link with bandwidth $B$ and signal-to-noise (SNR) $\rho$. According to the Shannon capacity, the maximum data transmission rate will be
\begin{equation}
C = B{\log _2}(1 + \rho )\
\end{equation}

If we only consider the transmission power $P_T$, then the EE will be
\begin{equation}
EE = \frac{C}{{{P_T}}}
\end{equation}

In [2], the relationship between SNR and EE is discussed by using the Shannon capacity formulation. As shown in Figure 1, it indicates that the minimum bit energy is achieved as $\rm{SNR} \to 0$, which is called low-SNR regime. But for a real communication system, low-SNR regime implies very low data rate and poor QoS. So we want to propose a more practical solution to achieve high EE performance.
\begin{figure}[!h]
\centering
%
%
\definecolor{mycolor1}{rgb}{0.00000,0.44700,0.74100}%
\begin{tikzpicture}[scale=0.7]

\begin{axis}[%
width=4.521in,
height=3.563in,
at={(0.758in,0.484in)},
scale only axis,
xmin=-20,
xmax=20,
xlabel style={font=\color{white!15!black}},
xlabel={$\rho$ (dB)},
ymin=0,
ymax=5*10^8,
ylabel style={font=\color{white!15!black}},
ylabel={EE (bit/J)},
axis background/.style={fill=white}
]
\addplot coordinates {
(-16, 494996069.090583)
(-12,	486010969.036091)
(-8,	465224313.578393)
(-4,	421890203.738297)
(0,	347396517.921382)
(4,	250635541.709491)
(8,	158006574.586491)
(12,	89311805.3199507)
(16, 46692847.5874997)
(20,	23130394.8469213)
};
\end{axis}

\end{tikzpicture}%
\vspace{-2mm}
\caption{EE performance with the change of SNR.}
\label{fig.1}
\end{figure}
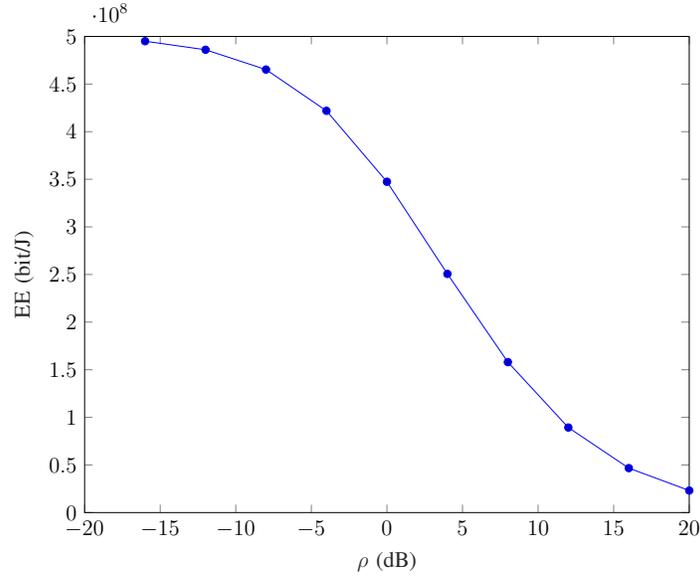

Fig. 1 considers the ideal case that power efficiency of radio amplifier is hundred percent and other circuit power consumption is ignored. Then, ${P_T}{\rm{ = }}\rho  \cdot B{N_0}$, where $N_0$ is the noise spectral density. From the figure, when the SNR $\rho$ is lower than nearly $-10$ dB, the EE maintain at a high level, but for $\rho>-10$ dB, the EE performance drops quickly until it gets nearly $20$ dB.

In the subsequent discussion, we assume that the cut-off value of SNR is $\rho_{th}$. Then the SNR is divided into two areas, where $\rho  \le {\rho _{th}}$ is called the high EE area and $\rho  > {\rho _{th}}$ is the decline area.

\subsection{Spacial Dimension and Frequency Dimension}
In this paper, we investigate how to keep the system working in the high EE area, which means the receiver SNR should be lower than the threshold. Consider a point-to-point MIMO system, where both the transmitter and the receiver are equipped with multiple antennas. The number of antennas is $M_s$ for the transmitter and $M_r$ for the receiver. If $\bm{H}$ is the channel matrix and $\bm{w}$ is the additive noise with a zero-mean circularly symmetric complex Gaussian distribution. Then, the received signal can be expressed as
\begin{equation}
\bm{y} = \sqrt {{\rho_D}} \bm{H}\bm{x} + \bm{w}
\end{equation}
where $\bm{x}$  represents the transmitted signal vector with normalized transmit power, $E \{ \Vert \bm{x} \Vert \}=1$. $\rho_D$ represents the corresponding SNR for transmission power $P_T$. To discuss the effect of resources on spacial dimension, we assume that all antennas on the transmitter are used, where $\bm{H} \in {C^{{M_s} \times {M_r}}}$. With the help of antennas, the transmission space can be divided into $M(M=min(M_s,M_r))$ parallel subchannels. Using the singular value decomposition (SVD) method, the channel matrix can be expressed as
\begin{equation}
{\bm{H}}{\rm{ = }}\bm{\phi} \sqrt {{\bm{D}}} \bm{\varphi}
\end{equation}
where $\bm{\phi}$ and $\bm{\varphi}$ are unitary matrices of dimension ${M_s} \times {M_s}$ and ${M_r} \times {M_r}$ respectively, and $\bm{D}$ is a block matric $\bm{D} = [{\bm{\Delta} _D},{\bm{0}_{M,\max ({M_s},{M_r}) - M}}]$, where ${\bm{\Delta} _D}{\rm{ = diag(}}{\lambda _1},{\lambda _2},...,{\lambda _M}{\rm{)}}$ denotes the singular values. If the power is uniformly allocated to all subchannels and perfect channel state information is known at the receiver, then the system's capacity is given by
\begin{equation}
{C_{MIMO}} = \sum\limits_{k = 1}^M {B_0}{{\log }_2}(1 + \frac{{{\rho _D}{\lambda _k}}}{{{M_s}}})
\end{equation}
where $B_0$ is the transmitted bandwidth.

From (5), the MIMO capacity is equivalent to the summation of $M$ parallel spacial subchannels with $\rm{SNR}=\frac{{{\rho _D}{\lambda _k}}}{{{M_s}}}$. Let $P_T$ to be the total transmission power of all the antennas, then EE for MIMO system will be
\begin{equation}
E{E_{MIMO}} = \frac{{{C_{MIMO}}}}{{{P_T}}}{\rm{ = }}\frac{{\sum\limits_{k = 1}^M {{B_0}{{\log }_2}(1 + \frac{{{\rho _D}{\lambda _k}}}{{{M_s}}}){\kern 1pt} } }}{{P_T}}
\end{equation}
On each subchannel, to ensure the system working in the high EE area, the received SNR should satisfy
\begin{equation}
\frac{{{\rho _D}{\lambda _k}}}{{{M_s}}} \le {\rho _{th}}
\end{equation}
From the formulation in (6) and (7) we can see that in spacial dimension, the multiple transmitted antennas are used to achieve low SNR communication and the $M$ subchannels are used to achieve high overall capacity.

Besides spacial dimension, we will further show that resources on frequency dimension has the similar effect. Considering for the system, only one transmitted antenna is used but the transmission bandwidth is adjustable. If the total bandwidth used is $B$, similar to the previous discussion, we denote the bandwidth into $N$ parallel frequency channels with same bandwidth $B_0=\frac{B}{N}$. The transmission power is also allocated to all frequency channels. Let $C_F$ be the total capacity, we can derive the EE of using frequency dimension as
\begin{equation}
E{E_F} = \frac{{{C_F}}}{{{P_T}}}{\rm{ = }}\frac{{\sum\limits_{i = 1}^N {{B_0}{{\log }_2}(1 + {\rho _i}{\gamma _i})} }}{{{P_T}}}
\end{equation}
and the constraint of high EE area for the SNR $\rho_i$ will be
\begin{equation}
{\rho _i}{\gamma _i} = \frac{{{p_i}{\gamma _i}}}{{{B_0}{N_0}}} \le {\rho _{th}}
\end{equation}
where ${\gamma _i}$ is the channel fading gain for frequency channel $i$. $p_i$ is the transmission power which $\sum\limits_{i = 1}^N {{p_i}}  = P_T$.

From the result in (8) and (9), frequency channels can be consider to replace the effect of multiple antennas to achieve both low SNR condition and high capacity need. There have been plenty of researches working on the EE analysis in spacial dimension or frequency dimension. Besides to analyze them separately, we decide to consider them as a unit to present a two-dimension resource allocation scheme based on the similarity when system is working in the high EE area.

\subsection{A Two-dimension Scheme}
We have presented the effect of both frequency and spacial dimensions to realize the communication in the high EE area and to achieve high capacity. But given the capacity formulation in (5), if the needed capacity is too high, the corresponding number when use one dimension resource also need to be sufficient large. In order to reduce the resource consumption for the single dimension, we decide to consider both frequency dimension and spacial dimension as a unit and present a new two-dimension scheme.

Fig.2 shows the schematic diagram of the scheme. From the figure, the new resource block is determined by subchannels in spacial dimension and frequency channels in frequency dimension, simultaneously. The number of frequency channels is $N$ and spacial subchannels is $M$. Resource block $R_{ij}$ represents the combination of frequency channel $i$ and spacial subchannel $j$. $\lambda_j^i$ is the channel coefficient and $p_{ij}$ is the transmission power. Then, the maximum transmission rate for $R_{ij}$ will be
\begin{equation}
{C_{{R_{ij}}}} = {B_0}{\log _2}(1 + {\rho _{ij}}{\lambda_j^i})
\end{equation}
where ${\rho _{ij}}{\rm{ = }}\frac{{{p_{ij}}}}{{{B_0}{N_0}}}$.

Under ideal conditions, we ignore the interaction between different resource blocks. Then, the maximum transmission rate using all resource blocks will be
\begin{equation}
{C_R} = \sum\limits_{i = 1}^N {\sum\limits_{j = 1}^M {{B_0}{{\log }_2}(1 + {\rho _{ij}}{\lambda_j^i})} }
\end{equation}

\begin{figure}[!h]
\centering
\includegraphics[width=4in]{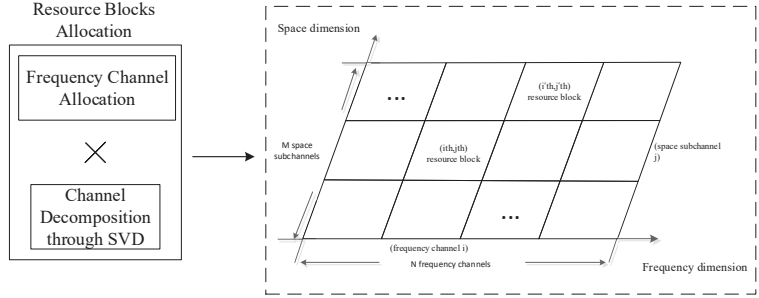}
\vspace{-2mm}
\caption{Resource blocks allocation progress.}
\label{fig:2}
\end{figure}

Then the EE with transmission power $P_T$ is
\begin{equation}
E{E_R} = \frac{{{C_R}}}{{{P_T}}}{\rm{ = }}\frac{{\sum\limits_{i = 1}^N {\sum\limits_{j = 1}^M {{B_0}{{\log }_2}(1 + \rho _{ij}{\lambda_j^i})} } }}{{{P_T}}}
\end{equation}

If the system is working in the high EE area, the SNR $\rho_{ij}$ for resource block $R_{ij}$ will subject to
\begin{equation}
\rho _{ij}{\rm{ = }}\frac{{{p_{ij}}}}{{{B_0}{N_0}}} \le {\rm {\rho}}_{th}
\end{equation}
and $\sum\limits_{j = 1}^M {\sum\limits_{i = 1}^N {{p_{ij}}}  = P_T}$.

Based on the discussion, the effect of the two-dimension schemes is seen in two aspects. First, according to the definition, the maximum number of resource blocks is $M \times N$. So, the average value for $p_{ij}$ will be $\frac{{{P_T}}}{{MN}}$. From the constraint in (13), it is more easier for the system to working in the high EE area. Another effect is for the system capacity. By taking two dimensions as a unit, the dynamically adjusting of resource on one dimension will lead to multiple growth on the resource blocks. This is the key to guarantee high capacity for the system. The detailed discussion will be given in the numerical part in Section IV.

dynamically adjusting the resources in different dimensions, the transmission rate is also guaranteed when communicating in the high EE area. The detailed discussion will be given in the numerical part in Section IV.

\section{Communication in The High EE Area with QoS Constraint}
We have presented a two-dimension scheme for the system to work in the high EE area. To use the scheme, the transmitter need to allocate resources in both frequency dimension and spacial dimension simultaneously to get the resource blocks. The system model is illustrated in Fig. 3. We concentrate on the discrete-time system over the point-to-point wireless link. It is assumed that data sequences are generated by the transmitter and the are divided into frames of duration $T$. In each frames, the data sequences are allocated to the two-dimension resource blocks. According to the constraint in (13), each resource block will work in the low SNR condition to improve the EE performance. Consdiering communication with low SNR may have bad effect on the QoS of the system\textsuperscript{\cite{chang2012performance}}. We also add QoS constraint into the power control progress.

\begin{figure}[!h]
\centering
\includegraphics[width=4in]{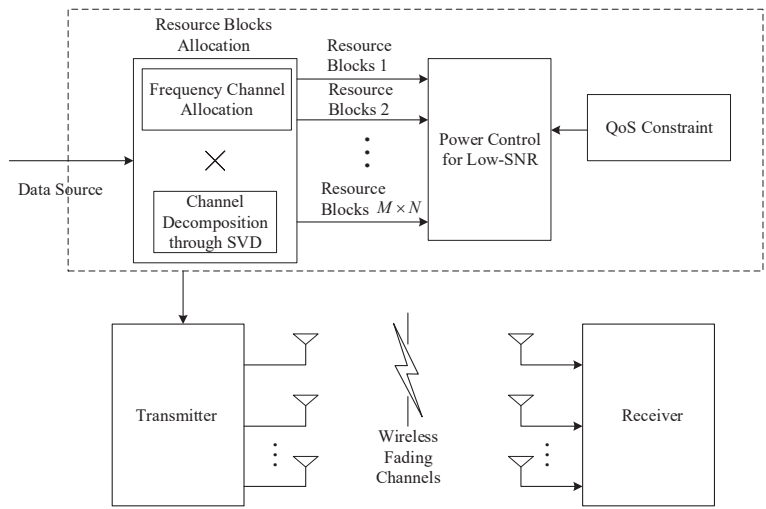}
\vspace{-2mm}
\caption{System model under QoS constraint.}
\label{fig:3}
\end{figure}
The system model is illustrated in Fig. 3. We concentrate on the discrete-time system over the point-to-point wireless link. It is assumed that data sequences are generated by the transmitter and the are divided into frames of duration $T$. We assume that $N$ frequency channels and $M$ spacial subchannels are used in the two-dimension scheme.
\subsection{Energy Efficiency Model}
With the QoS requirment, we will use the effective capacity which was proposed by Wu and Negi\textsuperscript{\cite{wu2003effective}} to analyze the corresponding EE model for the system. Effective capacity is defined as the maximum constant arrival rate that a given service process can support in order to guarantee a statistical QoS requirement specified by the QoS exponent $\theta  \ge 0$. If the effective capacity for resource block $R_{ij}$ is ${C_e}{(\theta )_{i,j}}$, then the final EE considering QoS requirement is expressed as
\begin{equation}
E{E_e} = \frac{{{C_{total}}(\theta )}}{{E\{ {P_T}\} }} = \frac{{\sum\limits_{i = 1}^N {\sum\limits_{j = 1}^M {{C_e}{{(\theta )}_{i,j}}} } }}{{E\{ {P_T}\} }}
\end{equation}
where ${E\{ {P_T}\} }$ is the expectation of the total transmission power, $N$ and $M$ are the number of used frequency channels and spacial subchannels, respectively. From the model expression, the EE for the system is affected by different QoS constraints exponent. Under ideal conditions, the fading process is assumed to be independent among frames and keep invariant within a frame duration. Then, the effective capacity ${C_e}{(\theta )_{i,j}}$ with a statistical QoS constraint exponent for resource block $R_{ij}$ can be expressed as follows\textsuperscript{\cite{tang2007quality}}
\begin{equation}
{C_e}{(\theta )_{i,j}} =  - \frac{1}{\theta }\log (E\{ {e^{ - \theta r_{ij}}}\} )
\end{equation}
where $r_{ij}$ denotes the instantaneous bit rate within a frame duration. According to (11), $r_{ij}$ can be expressed as
\begin{equation}
{r_{i,j}} = T{B_0}{\log _2}(1 + {\mu _{i,j}}(\theta ,\lambda ){\rho _{ij}}\lambda )
\end{equation}
where ${\mu _{i,j}}(\theta ,\lambda )$ is the power-adaptation factor which represents the transmission power allocated to frequency channel $i$ and spacial subchannel $j$. Substituting the bit rate into the EE model, we can get the final model expression as
\begin{equation}
E{E_e} = \frac{{{C_{total}}(\theta )}}{{E\{ {P_T}\} }} = \frac{{\sum\limits_{i = 1}^N {\sum\limits_{j = 1}^M { - \frac{1}{\theta }\log \left( {E\{ {e^{ - \theta T{B_0}{{\log }_2}(1 + {\mu _{i,j}}(\theta ,\lambda ){\rho _{ij}}\lambda )}}\} } \right)} } }}{{E\{ {P_T}\} }}
\end{equation}
When the system is working in the high EE area, the receiver SNR need to be lower than the SNR threshold. For $\rho_{ij}$, we have
\begin{equation}
{\rho _{ij}} \le {\rho _{th}}
\end{equation}
\subsection{Approximation in the High EE Area}
According to \cite{ge2014energy}, the expectation part in the effective capacity in (17) is
\begin{equation}
E\{ {e^{ - \theta {r_{i,j}}}}\} {\rm{ = }}\int\limits_0^\infty  {{e^{ - \theta T{B_0}{{\log }_2}(1 + {\mu _{i,j}}(\theta ,\lambda ){\rho _{ij}}\lambda )}}{p_{{\Gamma _{i,j}}}}(\lambda )d\lambda } )
\end{equation}
the effective capacity depends on the joint probability density function (pdf) of all the eigenvalues, denote by ${p_{{\Gamma _{i,j}}}}(\lambda )(i = 1,2,...,N,j = 1,2,...,M)$. According to \cite{ge2014energy}, which can be expressed as
\begin{equation}
\begin{aligned}
{p_{{\Gamma _{i,j}}}}(\lambda ) &= \underbrace {\int {...} }_{M - 1}\int {p({\lambda _1},{\lambda _2},...,{\lambda _M})} d{\lambda _m}d{\lambda _{m + 1}}...d{\lambda _k}\\
& (1 \le m < k \le M{\kern 1pt} {\kern 1pt} {\kern 1pt} and{\kern 1pt} {\kern 1pt} {\kern 1pt} m \ne i,k \ne i)
\end{aligned}
\end{equation}
where ${\lambda _1},{\lambda _2},...,{\lambda _M}$ are $M$ eigenvalues of a central Wishart matrix $\bm{\widehat W}$ which is defined by the coefficient matrix $\bm{H}_i$ and satisfy ${\lambda _1} \ge {\lambda _2} \ge ... \ge {\lambda _M}$. Then, giving the optimized power-adaptation factor $\mu _{i,j}^{opt}(\theta ,\lambda )$\textsuperscript{\cite{tang2007quality}} and the joint pdf, we have the final EE model as
\begin{small}
\begin{equation}
\hspace{-2mm}
E{E_e} = \frac{{\sum\limits_{i = 1}^N {\sum\limits_{j = 1}^M { - \frac{1}{\theta }\log \left( {\int_0^\infty  {\left( {{e^{ - \beta \ln (1 + \mu _{i,j}^{opt}(\theta ,\lambda ){\rho _{ij}}\lambda )}}} \right)}  \times \left( {\underbrace {\int {...} \int}_{M - 1} {p({\lambda _1},{\lambda _2},...,{\lambda _M})d{\lambda _m}d{\lambda _{m + 1}}...d{\lambda _k}} } \right) \times d\lambda } \right)} } }}{{E\{{P_T}\}}}
\end{equation}
\end{small}

Given the result in (21), when large number of resource blocks are used, how to deal with the complex pdf will be a great challenge. In order to solve the problem, we decide to provide an approximate performance formulation. If the system is working in the high EE area, the corresponding SNR should satisfy the low-SNR constraint. First, to simplify the EE model, we let $\rho_{th}$ to represent the SNR on each resource block so that we do not need to consider the low SNR constraint. Then, we will further derive a new approximate function to calculate the system's EE performance.

First, we will derive the limitation value of the EE performance considering the effective capacity. For the EE performance in (21), the following result provides the maximum value for certain $\theta$.
\begin{theorem}
For each resource block using in the two-dimension scheme, when $\rho_{th} \to 0$, the maximum EE performance with $P_T=0$ is
\begin{equation}
\mathop {\lim }\limits_{{\rho _{th}} \to 0} EE_e^{{R_{ij}}} = \mathop {\lim }\limits_{{\rho _{th}} \to 0} \frac{{{C_e}{{(\theta )}_{i,j}}}}{{{\rho _{th}}{B_0}{N_0}}} = \frac{{E\left\{ \lambda  \right\}}}{{{N_0}\ln 2}}
\end{equation}
where $E(\lambda)$ represents the expectation of the channel gain $\lambda$.
\end{theorem}

\emph{Proof}: See Appendix A.

With the limitation value in theorem 1, the EE performance $EE_e^{{R_{ij}}}$ for resource block $R_{ij}$ can be written as
\begin{equation}
EE_e^{{R_{ij}}} = \frac{{E\left\{ \lambda  \right\}}}{{{N_0}\ln 2}} - {\Delta _{ij}}({\rho _{th}})
\end{equation}
where ${\Delta _{ij}}({\rho _{th}})$ is a sublinear term which represents the performance gap between the limitation value and the actual value of the system. We consider that the system is working in the high EE area, which means the corresponding $\rho_{th}$ stay at low level. Based on that, we can derive an approximate expression for the sublinear term ${\Delta _{ij}}({\rho _{th}})$.
\begin{theorem}
For low SNR value $\rho_{th}$, the sublinear term in the EE performance in (29) can be expressed as
\begin{equation}
{\Delta _{ij}}({\rho _{th}}) \approx \frac{{{\rho _{th}}}}{{{N_0}\ln 2}}((\beta  + 1)E({\lambda ^2}) - \beta (E{(\lambda)^2})
\end{equation}
\end{theorem}

\emph{Proof}: See Appendix B.

Putting the approximate value of ${\Delta _{ij}}({\rho _{th}})$ into the EE performance formulation,we can get
\begin{equation}
EE_e^{{R_{ij}}} = \frac{{E\left\{ \lambda  \right\}}}{{{N_0}\ln 2}} - \frac{{{\rho _{th}}}}{{{N_0}\ln 2}}((\beta  + 1)E({\lambda ^2}) - \beta (E{(\lambda)^2})
\end{equation}
From the approximate expression, EE performance of each resource block is now related to the normalized QoS exponent $\beta$ and the statistical property of the channel gain $\lambda$. We successfully avoid the calculation of the complex pdf and achieve a simplify EE performance expression for the system working in the high EE area.

We will then consider the effect of the normalized QoS exponent $\beta$. For the formulation, the derivation of $\beta$ will be
\begin{equation}
\frac{{d(EE_e^{{R_{ij}}})}}{{d\beta }} =  - \frac{{{\rho _{th}}}}{{{N_0}\ln 2}}(E({\lambda ^2}) - (E{(z)^2}) =  - \frac{{{\rho _{th}}}}{{{N_0}\ln 2}}D(\lambda )
\end{equation}
where $D(\lambda )$ represents the variance of $\lambda$. For any random variable, the variance always has positive value. Then, the derivation will satisfies $\frac{{d(EE_e^{{R_{ij}}})}}{{d\beta }} < 0$. So with the increasing of $\beta$, the corresponding EE is monotone increasing. According to the definition given in \cite{tang2007quality}, large $\beta$ represents higher QoS requirements. That indicates that in order to meet QoS requirements, we need to sacrifices a part of system's EE performance. The result will be further analyzed in the numerical result part.

\section{Numerical Result}
In this section, we will give some numerical result about EE performance of the two-dimension scheme. We compare the scheme with a massive MIMO system. In the MIMO system, the numbers of antennas for the transmitter and receiver are ${M_s} = {M_r}{\rm{ = }}1024$. In the two-dimension scheme, the number of spacial subchannels is $M = 128$  and the number of frequency channels is $N = 100$. The bandwidth for each channel is ${B_0} = 10{\kern 1pt}$ Mhz. The transmission rate requirement for the system is increasing from ${10^8}$ bits/s to ${10^{10}}$ bits/s. We use the result in Section II to see the EE of both the systems.

As shown in the Fig.4, the horizontal line represents the transmission rate requirement for the system, and the vertical line is the corresponding EE. From the figure, with the increasing of transmission rate, the system using the two-dimension scheme keeps working in the high-EE area. But for MIMO system, the EE performance drops form the high EE area to the decline area when the transmission rate requirements increases. According to the numerical result, the effect of the two-dimension scheme is to keep working in the high EE area and to meet high requirements for the transmission rate.

\begin{figure}[!h]
\centering
%
%
\begin{tikzpicture}[scale=0.7]

\begin{axis}[%
width=4.521in,
height=3.563in,
at={(0.758in,0.484in)},
scale only axis,
xmode=log,
xmin=10000000,
xmax=1000000000,
xminorticks=true,
xlabel style={font=\color{white!15!black}},
xlabel={C(bit/s)},
ymin=695000000,
ymax=725000000,
ylabel style={font=\color{white!15!black}},
ylabel={EE(bit/J)},
axis background/.style={fill=white},
legend style={at={(0.157,0.149)}, anchor=south west, legend cell align=left, align=left, draw=white!15!black}
]
\addplot coordinates {
(10000000,	722815643.790543)
(15848931.9246111,	722672550.277182)
(25118864.3150958,	722445801.035637)
(39810717.0553497,	722086524.900381)
(63095734.4480193,	721517354.739785)
(100000000,	720615894.079094)
(158489319.246111,	719188715.629456)
(251188643.150958,	716930659.535325)
(398107170.553497,	713361600.882868)
(630957344.480193,	707729436.077312)
(1000000000,	698864368.484992)
};
\addlegendentry{massive MIMO for Ms=Mr=1024}

\addplot coordinates {
(10000000,	723021181.965473)
(15848931.9246111,	722998281.393144)
(25118864.3150958,	722961987.422432)
(39810717.0553497,	722904467.844428)
(63095734.4480193,	722813311.706106)
(100000000,	722668854.662618)
(158489319.246111,	722439945.11265)
(251188643.150958,	722077246.980975)
(398107170.553497,	721502657.997855)
(630957344.480193,	720592620.82869)
(1000000000,	719151879.034642)
};
\addlegendentry{the proposed scheme for M=64,N=100}

\end{axis}

\end{tikzpicture}%
\vspace{-2mm}
\caption{EE performance for Massive MIMO and the proposed scheme.}
\label{fig.4}
\end{figure}
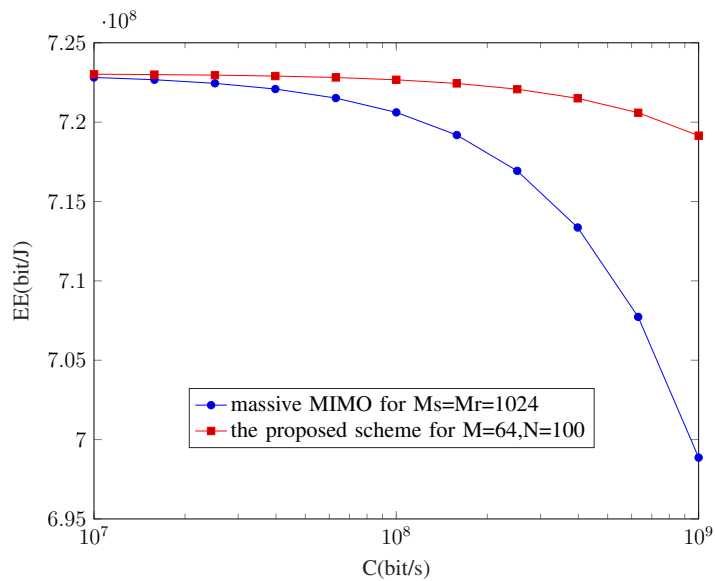

We further present a specific example to demonstrate the effect of our proposed scheme. Consider for a communication system, the needed transmission rate is $R = 5 \times {10^9}$ bit/s. For the system using one-dimension scheme, we will use the EE given in (8) to calculate the number of needed channels. For the system with the two-dimension scheme, we will use (13) to calculate the number of resource blocks. The result is shown in Table.1.

\begin{table}[h]
\centering
\caption{COMPARISON OF DIFFERENT SCHEMES.}
\label{TABLE I}
\begin{tabular}{{c c c c c c}}
\hline
\multirow{2}*{System Scheme}  & \multirow{2}*{$R$}  & \multirow{2}*{$\rho_{th}$} & \multicolumn{2}{c }{Resource} & \multirow{2}*{EE(J/bits)}    \\
\cline{4-5}
                              &      &            &  frequency    &     space  &   \\
\hline
One-dimension  & $ 5 \times {10^9}$  & $ - 10$ dB & 1200(channels) & 0 & $1.8 \times {10^{8}}$    \\
\hline
One-dimension  & $ 5 \times {10^9}$  & $ - 10$ dB & 0 & 1024(antennas) & $2 \times {10^{8}}$    \\
\hline
Two-dimension  & $ 5 \times {10^9}$  & $ - 10$ dB & 100(channels) & 64(channels) & $2.3 \times {10^{8}}$    \\
\hline
\end{tabular}
\end{table}

From TABLE I we can see, for the system using frequency channels, the required number to meet the transmission rate demand is $N \approx 1200$ and the corresponding EE is $E{E_F} = 1.8 \times {10^{8}}$. Similarly, for the massive MIMO system, the required number of antennas is ${M_s} = {M_r} =1024$ and the final EE can be expressed as $E{E_{MIMO}} = 2 \times {10^{8}}$. For the system with two-dimension scheme, the number of frequency channels is $N = 100$ and of spacial channels $M=64$ to achieve the same transmission rate. According to the result, one of the advantages for the two-dimension scheme is to reduce the resource consumption on each dimension.

Using the numerical result, we prove that the two-dimension scheme can help to keep the system working in high EE area. And to achieve the same transmission rate, the needed of each resource is several times lower comparing with the system with one-dimension cases. In the next part we will give some simulation results to evaluate the EE model when QoS constraint is included.

Fig.5 shows the EE for the system with the two-dimension scheme. In the simulation, the transmission power is $P_T=20$mw and the noise spectral density is $N_0=-174$dB/Hz. In the figure, we analyze the EE against both numbers of frequency channels and spacial channels. From the figure, both frequency dimension and spacial dimension have positive effect on EE for the system. But if the channels are used in the two-dimension scheme, the improvement of EE is more obvious. The result proves that the two-dimension scheme is the suitable way to realize the communication in the high EE area.
\begin{figure}[!h]
\centering
\input{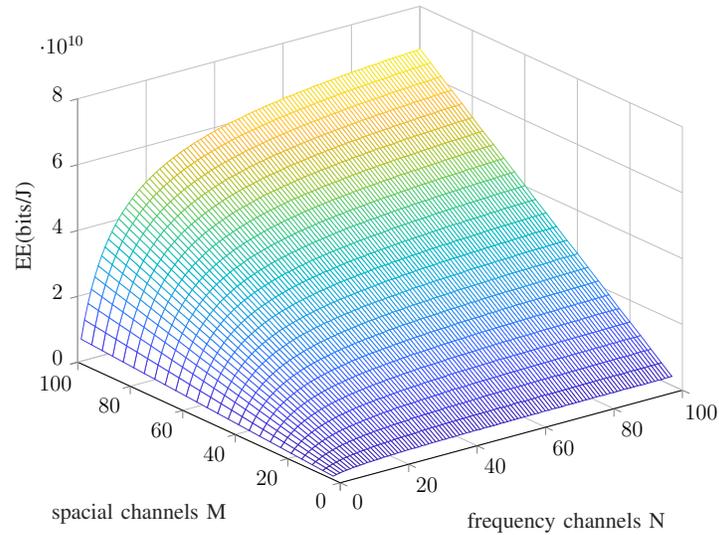}
\vspace{-2mm}
\caption{System's EE with the two-dimension scheme.}
\label{fig:5}
\end{figure}

In Fig.6, we shows the EE performance with the initial EE expression in (22) and the approximate expression in (27). From the figure we can see, with the corresponding SNR staying at low level, the EE calculated by both expressions are similar to each other. The result proves the feasibility of our approximate method to simplify the complex EE expression. With the simplified expression, we can easily analyze the effect of the QoS exponent.
\begin{figure}[!h]
\centering
%
%
\definecolor{mycolor1}{rgb}{0.00000,0.44700,0.74100}%
\definecolor{mycolor2}{rgb}{0.85000,0.32500,0.09800}%
\begin{tikzpicture}[scale=0.7]

\begin{axis}[%
width=4.521in,
height=3.538in,
at={(0.758in,0.509in)},
scale only axis,
xlabel style={font=\color{white!15!black}},
xlabel={$\text{Threshold SNR }\rho{}_{\text{th}}\text{ (dB)}$},
ymin=500000000,
ymax=750000000,
ylabel style={font=\color{white!15!black}},
ylabel={Energy Efficiency (bits/J)},
axis background/.style={fill=white},
legend style={at={(0.147,0.147)}, anchor=south west, legend cell align=left, align=left, draw=white!15!black}
]
\addplot coordinates {
(-28,	723184945.100222)
(-26,	721286795.964983)
(-24,	718313241.490464)
(-22,	713684990.248476)
(-20,	706551227.679374)
(-18,	695712962.014268)
(-16,	679582482.129866)
(-14,	656244842.279129)
(-12,	623699677.347313)
(-10,	580312282.450263)
};
\addlegendentry{EE Expression in (22)}

\addplot coordinates {
(-28,	720765083.120476)
(-26,	719422605.037431)
(-24,	717294920.662584)
(-22,	713922768.181183)
(-20,	708578266.669469)
(-18,	700107802.606457)
(-16,	686683021.776002)
(-14,	665406178.027531)
(-12,	631684653.213521)
(-10,	578239638.096389)
};
\addlegendentry{Approximate Expression in (27)}

\end{axis}

\begin{axis}[%
width=5.833in,
height=4.375in,
at={(0in,0in)},
scale only axis,
xmin=0,
xmax=1,
ymin=0,
ymax=1,
axis line style={draw=none},
ticks=none,
axis x line*=bottom,
axis y line*=left
]
\end{axis}
\end{tikzpicture}%
\vspace{-2mm}
\caption{Performance Analysis of the Approximate Expression with Low SNR.}
\label{fig.6}
\end{figure}
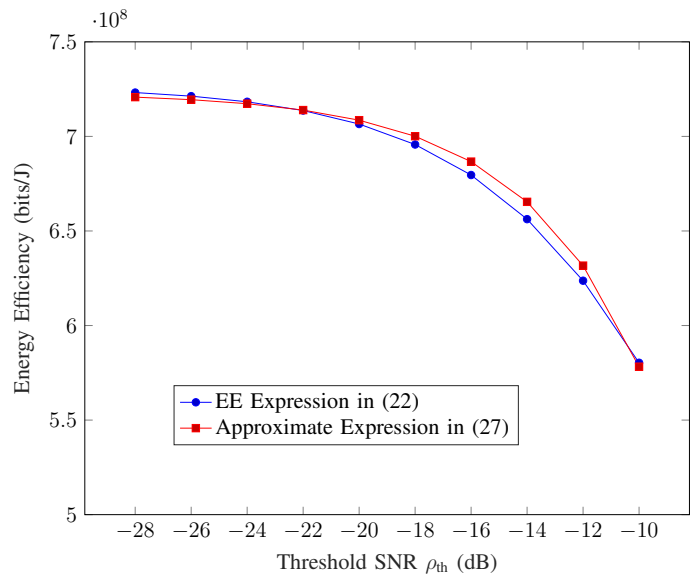

Fig.7 plots the EE as a function of the SNR threshold which is used to divide the high-EE area. We denote that resource blocks in the two-dimension scheme are used and the subchannel gain of the Rayleigh channel has $E\left\{ {\left| \lambda  \right|} \right\} = 1$. We compare different situations for different QoS exponent $\theta$. From the figure we notice that, with the increasing of the $\rho_{th}$, the EE still stay at high level. That means the system is working in the high EE area. Also form the figure, when QoS exponent $\theta$ is larger, the corresponding EE is lower and the decreasing trend of EE is more obvious. That prove the conclusion we derive from the approximate expression discussed in section III.

\begin{figure}[!h]
\centering
%
%
\begin{tikzpicture}[scale=0.7]

\begin{axis}[%
width=4.521in,
height=3.563in,
at={(0.758in,0.484in)},
scale only axis,
xmin=-20,
xmax=0,
xlabel style={font=\color{white!15!black}},
xlabel={$\text{Threshold value of SNR }\rho{}_{\text{th}}\text{ (db)}$},
ymin=15000,
ymax=50000,
ylabel style={font=\color{white!15!black}},
ylabel={Energy Efficiency (bits/J)},
axis background/.style={fill=white},
legend style={at={(0.728,0.78)}, anchor=south west, legend cell align=left, align=left, draw=white!15!black}
]
\addplot coordinates {
(-19,	44395.3433398367)
(-18,	44071.0038408724)
(-17,	43673.9225628575)
(-16,	43190.7826284731)
(-15,	42607.210467931)
(-14,	41908.3098028639)
(-13,	41079.4591640913)
(-12,	40107.3727301224)
(-11,	38981.3730976229)
(-10,	37694.7628302149)
(-9,	36246.1240172249)
(-8,	34640.3412000513)
(-7,	32889.1504040671)
(-6,	31011.072701723)
(-5,	29030.6859950378)
(-4,	26977.300443380)
(-3,	24883.2019061111)
(-2,	22781.6894377608)
(-1,	20705.145805047)
(0,	18683.3477827819)
};
\addlegendentry{$\theta\text{=1}$}

\addplot coordinates {
(-19,	45089.7080228169)
(-18,	44929.4479193071)
(-17,	44730.7818016114)
(-16,	44485.3946428624)
(-15,	44183.6184337345)
(-14,	43814.4271176972)
(-13,	43365.5439758213)
(-12,	42823.7041880773)
(-11,	42175.1121395882)
(-10,	41406.1191899672)
(-9,	40504.1206396488)
(-8,	39458.631205307)
(-7,	38262.4518766286)
(-6,	36912.7978895333)
(-5,	35412.2308365178)
(-4,	33769.2393118358)
(-3,	31998.3472736686)
(-2,	30119.6932520661)
(-1,	28158.1032008702)
(0,	26141.7566761204)
};
\addlegendentry{$\theta\text{-0.1}$}

\addplot coordinates {
(-19,	45160.7072282112)
(-18,	45017.687447027)
(-17,	44840.110539271)
(-16,	44620.345726074)
(-15,	44349.4433703588)
(-14,	44017.0854046648)
(-13,	43611.6256063217)
(-12,	43120.260401748)
(-11,	42529.3719573088)
(-10,	41825.0779508722)
(-9,	40994.0034481723)
(-8,	40024.2585295251)
(-7,	38906.563207555)
(-6,	37635.415912037)
(-5,	36210.1645709609)
(-4,	34635.8224767771)
(-3,	32923.4839874657)
(-2,	31090.2397469695)
(-1,	29158.5604511782)
(0,	27155.1975136394)
};
\addlegendentry{$\theta\text{=0.01}$}

\end{axis}

\begin{axis}[%
width=5.833in,
height=4.375in,
at={(0in,0in)},
scale only axis,
xmin=0,
xmax=1,
ymin=0,
ymax=1,
axis line style={draw=none},
ticks=none,
axis x line*=bottom,
axis y line*=left
]
\end{axis}
\end{tikzpicture}%
\vspace{-2mm}
\caption{EE performance for resource block under different QoS constraint.}
\label{fig.7}
\end{figure}
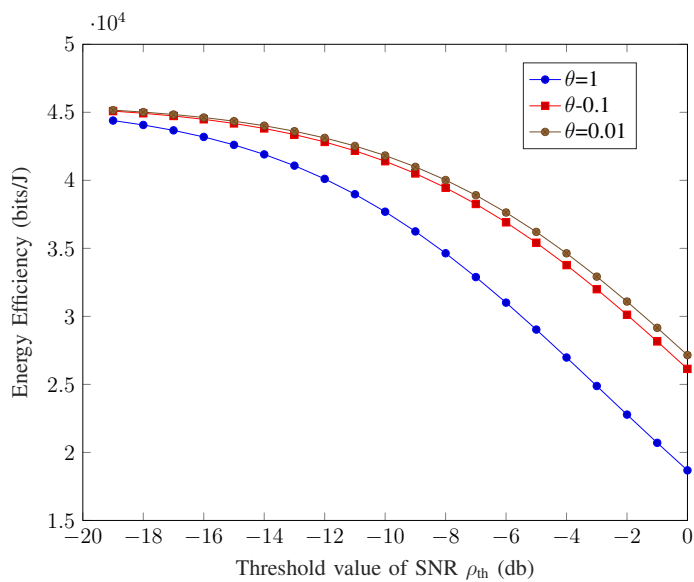

Fig.8 and Fig.9 shows the EE and the effective capacity of MIMO systems when two-dimension scheme is used. The number of transmitter antennas and receiver antennas are different for different system. In the figure, the QoS exponent is settled as $\theta=0.01$ and threshold value $\rho_{th}=-10$ dB. We plot the EE and effective capacity versus the transmitted bandwidth to see the effect of the dynamic change on frequency dimension. From the Fig.8 the EE show that all the systems are working in the high EE area. And according to the result in Fig.9, with the increasing of transmitted bandwidth, there is a significant increase in system effective capacity. That indicates the purpose of the scheme which is to dynamic change frequency resources to meet the high transmission rate requirements in future communication system. We also compare the performance with different antenna numbers. In the figure, for the effective capacity, the number of the receiver antennas $M_r$ is more effective comparing with transmitter antennas numner $M_s$. That is because the subchannel number $M$ is regarded as $\min \{ {M_s},{M_r}{\rm{\} }}$. Fot the EE, since the transmitted power is related to the number of transmitter antennas, large $M_s$ will cause a slight decrease.

\begin{figure}[!h]
\centering
%
%
\begin{tikzpicture}[scale=0.7]

\begin{axis}[%
width=4.521in,
height=3.563in,
at={(0.758in,0.484in)},
scale only axis,
xmode=log,
xmin=10000000,
xmax=1000000000,
xminorticks=true,
xlabel style={font=\color{white!15!black}},
xlabel={Transmitted Bandwidth B (Hz)},
ymin=0,
ymax=800000,
ylabel style={font=\color{white!15!black}},
ylabel={Energy Efficiency (bits/J)},
axis background/.style={fill=white},
legend style={at={(0.134,0.756)}, anchor=south west, legend cell align=left, align=left, draw=white!15!black}
]
\addplot coordinates{
(10000000,	72041.0672749883)
(15848931.9246111,	92244.7274362596)
(25118864.3150958,	113148.122650275)
(39810717.0553497,	129679.829498552)
(63095734.4480193,	145565.838288155)
(100000000,	158592.322302554)
(158489319.246111,	167315.312457799)
(251188643.150958,	172336.959043095)
(398107170.553497,	176381.774230132)
(630957344.480193,	181577.453464622)
(1000000000,	179199.957770353)
};
\addlegendentry{Ms=64,Mr=4}

\addplot coordinates{
(10000000,	280580.869336097)
(15848931.9246111,	363154.451191154)
(25118864.3150958,	438932.782118045)
(39810717.0553497,	521326.97352781)
(63095734.4480193,	575619.193848059)
(100000000,	621262.611809327)
(158489319.246111,	648164.395655589)
(251188643.150958,	676263.323284785)
(398107170.553497,	687145.096815418)
(630957344.480193,	700818.651441473)
(1000000000,	711712.776779737)
};
\addlegendentry{Ms=64,Mr=16}

\addplot coordinates{
(10000000,	57867.7577238396)
(15848931.9246111,	68151.9365570176)
(25118864.3150958,	79075.6438645148)
(39810717.0553497,	85540.3380966747)
(63095734.4480193,	91408.3592326167)
(100000000,	95747.2262459077)
(158489319.246111,	97785.6308847446)
(251188643.150958,	100587.725618668)
(398107170.553497,	100623.925473538)
(630957344.480193,	102657.459483279)
(1000000000,	102075.3296819)
};
\addlegendentry{Ms=128,Mr=4}

\addplot coordinates{
(10000000,	227747.193073486)
(15848931.9246111,	273501.994477607)
(25118864.3150958,	311821.517201173)
(39810717.0553497,	340800.385078407)
(63095734.4480193,	362335.529248144)
(100000000,	378147.413833086)
(158489319.246111,	389634.972044453)
(251188643.150958,	395351.878478914)
(398107170.553497,	400467.992238932)
(630957344.480193,	401042.440906636)
(1000000000,	406895.432241727)
};
\addlegendentry{Ms=128,Mr=16}

\end{axis}

\begin{axis}[%
width=5.833in,
height=4.375in,
at={(0in,0in)},
scale only axis,
xmin=0,
xmax=1,
ymin=0,
ymax=1,
axis line style={draw=none},
ticks=none,
axis x line*=bottom,
axis y line*=left
]
\end{axis}
\end{tikzpicture}%
\vspace{-2mm}
\caption{Energy efficiency versus transmitted bandwidth using two-dimension scheme.}
\label{fig:8}
\end{figure}
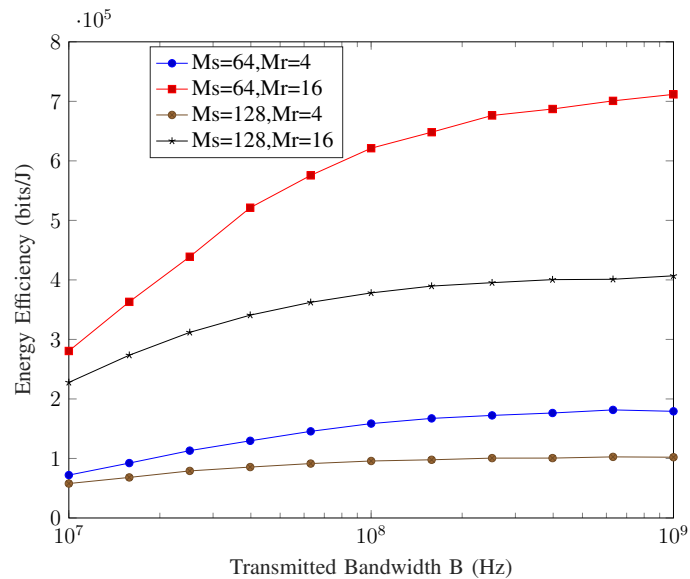

\begin{figure}[!h]
\centering
%
%
\begin{tikzpicture}[scale=0.7]

\begin{axis}[%
width=4.521in,
height=3.563in,
at={(0.758in,0.484in)},
scale only axis,
xmode=log,
xmin=10000000,
xmax=1000000000,
xminorticks=true,
xlabel style={font=\color{white!15!black}},
xlabel={Transmitted Bandwidth B (Hz)},
ymin=0,
ymax=12000000,
ylabel style={font=\color{white!15!black}},
ylabel={Effective Capacity (bits)},
axis background/.style={fill=white},
legend style={at={(0.153,0.75)}, anchor=south west, legend cell align=left, align=left, draw=white!15!black}
]
\addplot coordinates{
(10000000,	23407.223949405)
(15848931.9246111,	37405.3552150784)
(25118864.3150958,	56908.1879246178)
(39810717.0553497,	94978.8092277866)
(63095734.4480193,	145377.313948458)
(100000000,	234533.032903912)
(158489319.246111,	371459.626348429)
(251188643.150958,	587048.535937817)
(398107170.553497,	928899.240418139)
(630957344.480193,	1489204.83351063)
(1000000000,	2372456.09011377)
};
\addlegendentry{Ms=64,Mr=4}

\addplot coordinates{
(10000000,	91610.4493651672)
(15848931.9246111,	145594.988470717)
(25118864.3150958,	233352.292702557)
(39810717.0553497,	365681.408071177)
(63095734.4480193,	582113.35486689)
(100000000,	923676.006420346)
(158489319.246111,	1457005.96472772)
(251188643.150958,	2302952.13230507)
(398107170.553497,	3652193.5876567)
(630957344.480193,	5827347.96042003)
(1000000000,	9255672.74937138)
};
\addlegendentry{Ms=64,Mr=16}

\addplot coordinates{
(10000000,	26368.4728378452)
(15848931.9246111,	41940.6439505152)
(25118864.3150958,	67056.0344653192)
(39810717.0553497,	103576.218975103)
(63095734.4480193,	166420.943383766)
(100000000,	261311.34002603)
(158489319.246111,	418414.365823183)
(251188643.150958,	658533.979767738)
(398107170.553497,	1051137.3743674)
(630957344.480193,	1641057.04380854)
(1000000000,	2668550.60349353)
};
\addlegendentry{Ms=128,Mr=4}

\addplot coordinates{
(10000000,	104814.320224529)
(15848931.9246111,	165731.613374162)
(25118864.3150958,	262002.480507745)
(39810717.0553497,	414381.85317169)
(63095734.4480193,	655668.509338162)
(100000000,	1047245.62975671)
(158489319.246111,	1642610.97591716)
(251188643.150958,	2609724.49860904)
(398107170.553497,	4161122.16134835)
(630957344.480193,	6610082.81960346)
(1000000000,	10457625.6119269)
};
\addlegendentry{Ms=128,Mr=16}

\end{axis}

\begin{axis}[%
width=5.833in,
height=4.375in,
at={(0in,0in)},
scale only axis,
xmin=0,
xmax=1,
ymin=0,
ymax=1,
axis line style={draw=none},
ticks=none,
axis x line*=bottom,
axis y line*=left
]
\end{axis}
\end{tikzpicture}%
\vspace{-2mm}
\caption{Effective capacity versus transmitted bandwidth using two-dimension scheme.}
\label{fig:9}
\end{figure}
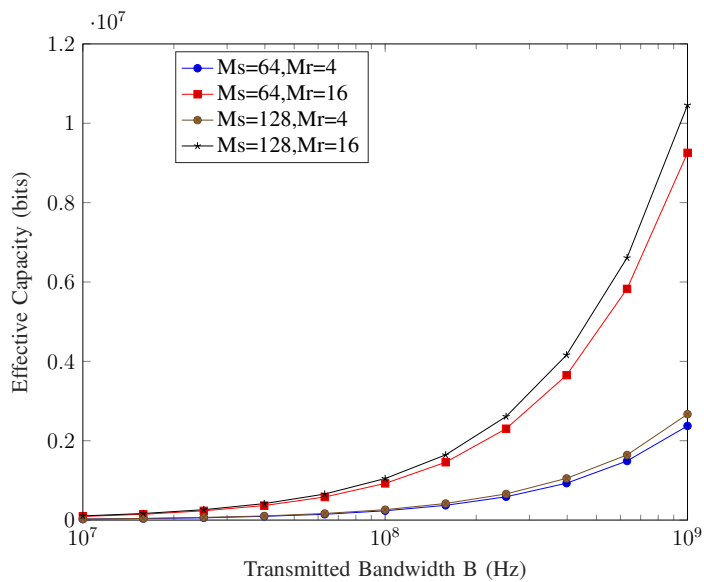

\section{Conclusion}
In this paper, we have proposed a new two-dimension scheme aiming at achieving high EE communication. To this end, we presented an idea of high EE area which is defined by receiver SNR. We have discussed the EE in spacial and frequency dimension and find equivalence between them. Based on that, we further decided to consider them as a unit and presented a two-dimension scheme. In the scheme, each resource block is determined by both frequency channels and spacial channels simultaneously. With the resource blocks, the system can realize the high EE communication and meanwhile to meet increasing capacity requirements. Moreover, we have derived the EE model for the system when QoS constraint and effective capacity are under consideration. Since the model is too complex to analyse, we use the low SNR condition to get an approximate result. Finally, the numerical result have been presented to verify the conclusion.

\begin{appendices}
\section{}
According to \cite{verdu2002spectral}, for ${\rho _{th}} \to 0$, the bit energy ${E_b^{{R_{ij}}}}$ for resource block $R_{ij}$ can be defined as
\begin{equation}
\mathop {\lim }\limits_{{\rho _{th}} \to 0} \frac{{E_b^{{R_{ij}}}}}{{{n_0}}} = \frac{{T{B_0}}}{{{C_e}{{(\theta )}_{i,j}}^\prime }}
\end{equation}
So, the EE for the resource block is
\begin{equation}
\mathop {\lim }\limits_{{\rho _{th}} \to 0} EE_e^{{R_{ij}}} = \frac{{{C_e}{{(\theta )}_{i,j}}^\prime }}{{TB{N_0}}}
\end{equation}
Considering the formulation in (18), let $F({\rho _{th}}) = E({e^{ - \beta \ln (1 + \mu \left( {\theta ,\lambda } \right){\rho _{th}}\lambda )}})$, then we can derive the derivative as
\begin{equation}
{C_e}{(\theta )_{i,j}}^\prime  =  - \frac{1}{\theta } \cdot \frac{1}{{F({\rho _{th}})}} \cdot \frac{{dF({\rho _{th}})}}{{d{\rho _{th}}}}
\end{equation}
For the derivative of ${F({\rho _{th}})}$, we have
\begin{equation}
\frac{{dF({\rho _{th}})}}{{d{\rho _{th}}}} = E({(1 + \mu \left( {\theta ,\lambda } \right){\rho _{th}}\lambda )^{ - \beta  - 1}}\lambda )
\end{equation}
Put the result into (30), we have
\begin{equation}
{C_e}{(\theta )_{i,j}}^\prime  =  - \frac{1}{\theta } \cdot \frac{{E({{(1 + \mu \left( {\theta ,\lambda } \right){\rho _{th}}\lambda )}^{ - \beta  - 1}}\lambda )}}{{E({e^{ - \beta \ln (1 + \mu \left( {\theta ,\lambda } \right){\rho _{th}}\lambda )}})}}
\end{equation}
Let the ${\rho _{th}} \to 0$ then we can get $\mathop {\lim }\limits_{{\rho _{th}} \to 0} {C_e}{(\theta )_{i,j}}^\prime  = \frac{{TB}}{{\ln 2}} \cdot E(\lambda )$. Putting the result in (23), we can get the result in theorem 1.

\section{}
For a certain $\theta$, we use $f({\rho _{th}})$ to represents the ${C_e}{(\theta )_{i,j}}$ in the EE performance formulation. Then, we can derive the Maclaurin's expansion of it as
\begin{equation}
f({\rho _{th}}) = f(0) + f'(0){\rho _{th}} + f''(0)\rho _{th}^2 + o(\rho _{th}^2)
\end{equation}
where $o(\rho _{th}^2)$ represents the high order term for $rho_{th}$. In the high EE area, $\rho_{th}$ satisfies the low SNR constraint. So the the high order term can be ignored. In Appendix A, we derive the first derivation of $f({\rho _{th}})$ when $\rho_{th}=0$. Similarity, we can further derive the second order of $f({\rho _{th}})$ as
\begin{equation}
f''(0) = \frac{1}{{\ln 2}}((\beta  + 1)E({\lambda ^2}) - \beta (E{(\lambda)^2})
\end{equation}
Putting the result in (33), we can get the approximate formulation as
\begin{equation}
f({\rho _{th}}) = \frac{{E\left\{ \lambda  \right\}}}{{\ln 2}} \cdot {\rho _{th}} - \frac{1}{{\ln 2}}((\beta  + 1)E({\lambda ^2}) - \beta (E{(z)^2}) \cdot \rho _{th}^2
\end{equation}
Putting $f({\rho _{th}})$ into the EE performance formulation, we can get the approximate equation in theorem 2.
\end{appendices}

\bibliographystyle{IEEEtran}
\bibliography{mybibtex.bib}

\end{document}